\documentclass[journal=cmatex, manuscript=article, layout=twocolumn]{achemso}

\usepackage[utf8]{inputenc}
\usepackage[english]{babel}
\usepackage{mhchem}
\usepackage{graphicx}
\usepackage{subcaption}

\author{Niek J.J. de Klerk}
\affiliation[Delft]{Department of Radiation Science and Technology, Delft University of Technology, Mekelweg 15, 2629JB Delft, The Netherlands}
\author{Eveline van der Maas}
\affiliation[Delft]{Department of Radiation Science and Technology, Delft University of Technology, Mekelweg 15, 2629JB Delft, The Netherlands}
\author{Marnix Wagemaker}
\affiliation[Delft]{Department of Radiation Science and Technology, Delft University of Technology, Mekelweg 15, 2629JB Delft, The Netherlands}
\email{m.wagemaker@tudelft.nl}

\title{Analysis of diffusion in solid state electrolytes through MD-simulations, improvement of the Li-ion conductivity in \ce{\beta-Li3PS4} as an example}

\begin{document}
\maketitle

\begin{abstract}
Molecular dynamics simulations are a powerful tool to study diffusion processes in battery electrolyte and electrode materials. From a single molecular dynamics simulation many properties relevant to diffusion can be obtained, including the diffusion path, attempt frequency, activation energies, and collective diffusion processes. These detailed diffusion properties provide a thorough understanding of diffusion in solid electrolytes, and provides direction for the design of improved solid electrolyte materials.  
Here a thorough analysis methodology is developed, and applied to DFT MD simulations of Li-ion diffusion in \ce{\beta-Li3PS4}. The methodology presented is generally applicable to crystalline materials and facilitates the analysis of molecular dynamics simulations. The code used for the analysis is freely available at: \\ https://bitbucket.org/niekdeklerk/md-analysis-with-matlab. \\
The results on \ce{\beta-Li3PS4} demonstrate that jumps between bc-planes limit the conductivity of this important class of solid electrolyte materials.
The simulations indicate that by adding Li-interstitials or Li-vacancies the rate limiting jump process can be accelerated significantly, which induces three dimensional diffusion, resulting in an increased Li-ion diffusivity. 
Li-vacancies can be introduced through Br-doping, which is predicted to result in an order of magnitude larger Li-ion conductivity in \ce{\beta-Li3PS4}. Furthermore, the present simulations rationalise the improved Li-ion diffusivity upon O-doping through the creation of local Li-interstitials. 
\end{abstract}

\section{Introduction}  
To prevent further global warming by greenhouse gas emissions it is necessary to move from fossil fuels towards renewable energy sources. For transport applications other energy carriers, such as hydrogen and batteries, are considered. Of the current technologies which can replace fossil fuels in vehicles, batteries result in the lowest greenhouse gas emissions \cite{Bauer_2015}, especially if renewable sources are used for the energy production. \\ 
However, safety concerns and the limited range of current battery electric vehicles are slowing down their implementation. Solid state batteries are a promising technology \cite{Lotsch_2017, Placke_2017} based on the much lower flammability risks, the higher energy density on the cell level, and lower self-discharge rate. \\
One of the prerequisites towards the realisation of solid state batteries is the development of highly conductive solid electrolytes. In recent years several materials have been discovered which show conductivities comparable to liquid electrolytes. Room temperature ionic conductivities in the order of $10^{-3}$ S/cm have been reported in a range of lithium containing compounds \cite{Lotsch_2017}, such as LLTO, argyrodites, LGPS, and LATP. Fewer sodium containing compounds with such high conductivities are known, most likely the result of less intensive research in this area, but several have been established, including $\beta$-alumina \cite{Lu_2010} and \ce{Na3PS4} \cite{Chu_2016}. 
The combination of high ionic conductivity and large electrochemical stability is challenging \cite{Lotsch_2017, Zhu_2015}, but electrochemical stability does not have very strict requirements. A solid electrolyte can be successful if its decomposition products are stable, have a reasonable ionic conductivity and low electronic conductivity \cite{Zhu_2015}, similar to the functioning of solid-electrolyte interface (SEI) layers for liquid electrolytes. 
The complex demands on solid electrolytes necessitates fundamental research towards solid electrolyte properties and new solid electrolyte materials. \\
Computer simulations are playing an important role in understanding and directing materials design towards improved battery performance. For example, calculations have shown that the electrochemical stability of solid electrolytes is enhanced by passivating decomposition products \cite{Zhu_2015}, that strain can enhance diffusion \cite{Tealdi_2016}, how Li-ion diffusion can be increased in anti-perovskites \cite{Deng_2015_perov} and \ce{Na3PS4} \cite{Chu_2016, Klerk_Na}, and why bond frustration is beneficial for Li-ion diffusion \cite{Adelstein_2016}. \\
In solid state electrolytes the high concentration of diffusing atoms, 31 mol/L in \ce{\beta-Li3PS4}, can lead to complex interactions and diffusion behaviour. Diffusion can involve collective jumps \cite{Alex_hollandite} and lattice vibrations \cite{Klerk_Na, Phani_2017}, which are all included in molecular dynamics (MD) simulations since all possible motions of ions and their interactions are taken into account.
Furthermore, MD simulations can show unanticipated diffusion behaviour \cite{Yang_2011}, whereas static calculations (e.g. nudged elastic band and bond valence) are limited by the imagination of the researcher.
To understand diffusion in solid state electrolytes MD simulations are thus a powerful tool, allowing the dynamic diffusion processes to be studied in detail. \\ 
Although MD simulations have been shown to provide understanding of complicated diffusion processes, typically only the tracer diffusivity is extracted, from which an activation energy is calculated by assuming Arrhenius behaviour. A thorough analysis of MD simulations is able to give much more detailed results, potentially providing more understanding and concrete direction towards the design of improved conductivities \cite{Klerk_Li, Klerk_Na, Ganapathy_LTO}.      
In order to make such thorough analysis of MD simulations more easily available we present an approach, here demonstrated for \ce{\beta-Li3PS4}, that allows to extract the detailed diffusional properties based on a MD simulation and the crystalline structure of the studied material. The approach determines jump rates, activation energies, attempt frequency, vibrational amplitude, radial distribution functions, possible collective motions, site occupancies, tracer diffusivity, and the correlation factor. \\
The first part of this paper describes the approach that is followed to obtain the diffusional properties from a single MD simulation. In the second part MD simulations on \ce{\beta-Li3PS4} are analysed, exemplifying how the developed approach helps in understanding diffusion in solid state electrolytes, and how this provides direction to design improved solid electrolyte materials. The Matlab code used for the analysis of MD simulations is freely available online \cite{Bitbucket}. 

\section{Information from MD simulations}
After performing a MD simulation the position of all the atoms at every time step is known. Typically, this result is used to determine the tracer diffusivity ($D^{*}$) via the mean squared displacement \cite{Friauf_1962}: 
\begin{equation}
	D^{*} = \frac{1}{2 d N} \sum_{i=1}^{N} \left( \frac{r_{i}(t)}{t}^2 \right)
	\label{eq:d_msd}
\end{equation}
where $r_{i}(t)$ is the displacement of a single atom, $t$ the simulated time, $N$ the number of diffusing atoms, and $d$ the number of diffusion dimensions. Using the diffusivity and the Nernst-Einstein relation the conductivity ($\sigma$) can be determined \cite{Friauf_1962}:  
\begin{equation}
	\sigma = \frac{n e^2 z^2}{k_B T} D^{*}
	\label{eq:conductivity}
\end{equation}
where $n$ is the diffusing particle density, $e$ the elementary electron charge, $z$ the ionic charge, $k_B$ Boltzmann's constant, and $T$ the temperature in Kelvin.
Provided that the atomic displacement is significantly larger than the vibration amplitude, the tracer diffusivity provides a good quantitative indication of the ionic diffusion. 
However, to get a thorough understanding much more properties related to the diffusion process can be obtained from a single MD simulation, including: 
\begin{itemize} 
	\item Amplitude of vibrations
	\item Attempt frequency
	\item Site occupations
	\item Jump rates
	\item Correlation factor
	\item Activation energies
	\item Collective jumps
	\item Radial Distribution Functions
\end{itemize}

\subsection{Amplitude of vibrations}
\begin{figure}[htbp]
	\begin{center}
		\includegraphics[width=0.45\textwidth]{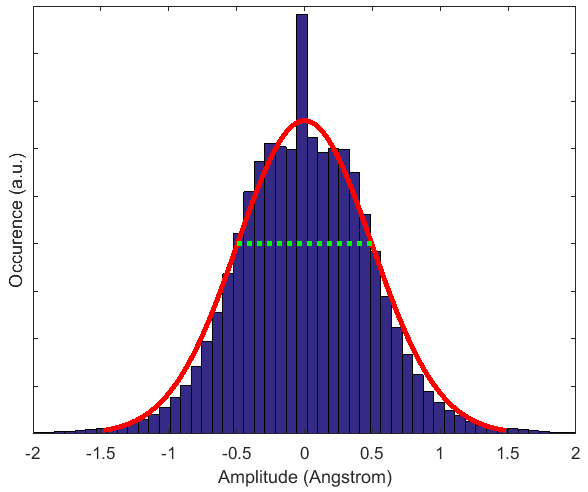}	
		\caption{Histogram showing the vibrational amplitude of Li-ions in \ce{\beta-Li3PS4} at 600 K, with the fitted Gaussian (solid red line) and the standard deviation ($\pm$ 0.495 \AA, dotted green line)}
		\label{fig:vib_amplitude_distr}
	\end{center}
\end{figure}

Atomic vibrations in a crystal are the 'back and forth' movement of an atom around a (meta)stable position. 
From a MD simulation the position of all atoms is known at any time, hence the direction of movement and displacement can be obtained.  
The atomic vibrations can be determined by monitoring the change in the direction of movement, and the vibrational amplitude is obtained by integrating the displacement of each atom until the derivative of the displacement changes sign, which corresponds to a change in the direction of movement. Doing this for all the atoms of interest gives a distribution of vibrational displacements, an example of which is shown in Fig. \ref{fig:vib_amplitude_distr}. 
By fitting a Gaussian function to the obtained distribution the standard deviation in vibrational displacement can be obtained, providing an estimate of the average amplitude of vibrations in the crystal. 
The obtained average amplitude of vibrations can be used as an estimate for (the temperature dependence of) the Debye-Waller factor. Since the 3D distribution is known, anisotropic vibrational amplitudes can also be determined from a MD simulation in this way.  

\subsection{Attempt frequency}
\begin{figure}[htbp]
	\begin{center}
		\includegraphics[width=0.45\textwidth]{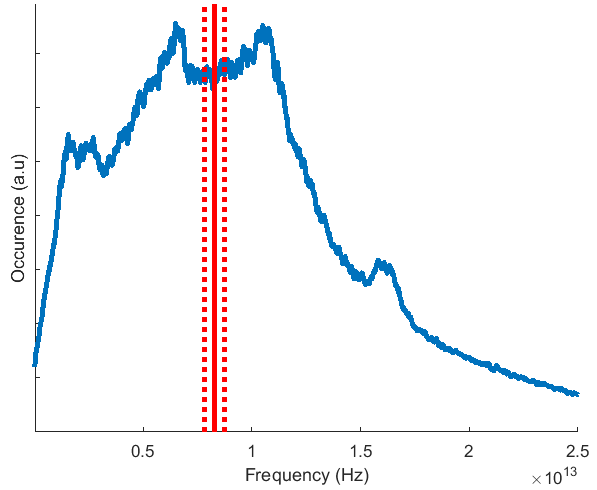}	
		\caption{Vibration frequency spectrum of Li-ions in \ce{\beta-Li3PS4} at 600 K, the average frequency of $8.29 (\pm 0.46) *10^{12} $ Hz is shown by the solid ($\pm$ dotted) red line.}
		\label{fig:freq_distr}
	\end{center}
\end{figure}

Based on (the sign of) the derivative of displacement the vibration time of an atom is known, from which the vibrational spectrum can be obtained via a Fourier transformation, shown for Li-ions in \ce{\beta-Li3PS4} in Fig. \ref{fig:freq_distr}. 
From the frequency spectrum the attempt frequency can be obtained if we consider every vibration of a (diffusing) atom as an attempt. Thus we define the average vibration frequency as the attempt frequency ($\nu^{*}$), which is necessary to determine the relation between jump rates and activation energies. \\
The approach of obtaining the attempt frequency presented here is very different from the usual approaches. Several definitions of the attempt frequency exist \cite{Koettgen_2017}, all of which require the determination of the transition state and calculation of the phonon spectrum for the stable and the transition state \cite{Versteylen_2017}. However, often these calculations are not performed, and a 'standard value' of $1*10^{13}$ Hz is used \cite{vdVen_2001, Koettgen_2017}. \\
In comparison to other methods the present method is straightforward, using the information that is already present in a MD simulation.  
Furthermore, since the attempt frequency is obtained from a single MD simulation the influence of temperature, structural parameters, etc., on the attempt frequency are included and can be investigated.

\subsection{Site occupancy}
In crystalline ionic conductors diffusion occurs through transitions between relatively stable sites. Typically these crystallographic sites are known from diffraction experiments, but if these are not known the sites can be extracted from a MD simulation through data mining \cite{Chen_2017}.
The condition used for site occupancy is that the distance of the ion to the centre of the crystallographic site is smaller than the site-radius.  
At present the site-radius is defined as twice the vibrational amplitude, which can for instance be reduced to prevent sites from overlapping.

\subsection{Jump rates}
When the crystallographic sites are known, detecting the transitions between sites that occur in a MD simulation is straightforward. 
Counting the number of jumps ($J$) between (types of) sites provides the mean jump rate ($\Gamma$) using: 
\begin{equation}
	\Gamma = \frac{J}{Nt}
	\label{eq:gamma}
\end{equation}
where $N$ is the number of diffusing atoms, and $t$ the simulation time. 
The standard deviation in the jump rate can be calculated by dividing the simulation into several parts, in order to estimate the reliability of the MD simulations.
As demonstrated recently \cite{Klerk_Li} determination of the different jump rates in a crystal provides direct insight in which jump process is rate-limiting for diffusion. This information can be used to design crystal structures with larger atomic diffusivity. 
Because NMR relaxation experiments can directly probe the jump rates, comparison with the jump rates from MD simulations can be used to validate the MD simulations \cite{Yu_Na_dynamics}, or to better understand the complex results from NMR experiments \cite{Yu_arg}. \\
Using the Einstein-Smulochowski relation the jump rates are related to the jump rate diffusivity ($D_{J}$):  
\begin{equation}
	D_{J} =  \sum_{i}^{} \frac{\Gamma a_i^2}{2 d}
	\label{eq:d_j}
\end{equation}
where $i$ are the different types of jumps, $a_i$ is the jump distance of jump type $i$, and $d$ the number of diffusion dimensions. \\
To get an estimate of the fraction of jumps which contribute to macroscopic diffusion the correlation factor ($f$), also known as the Haven ratio ($H_r$), is calculated \cite{Uitz_2017, vdVen_2001}:  
\begin{equation}
	f = \frac{D^*}{D_J}
	\label{eq:corr_factor}
\end{equation}

\subsection{Activation energies}
The probability that an actual jump from one site to another occurs is determined by the activation energy for this transition. 
\begin{figure}[htbp]
	\begin{center}
		\includegraphics[width=0.45\textwidth]{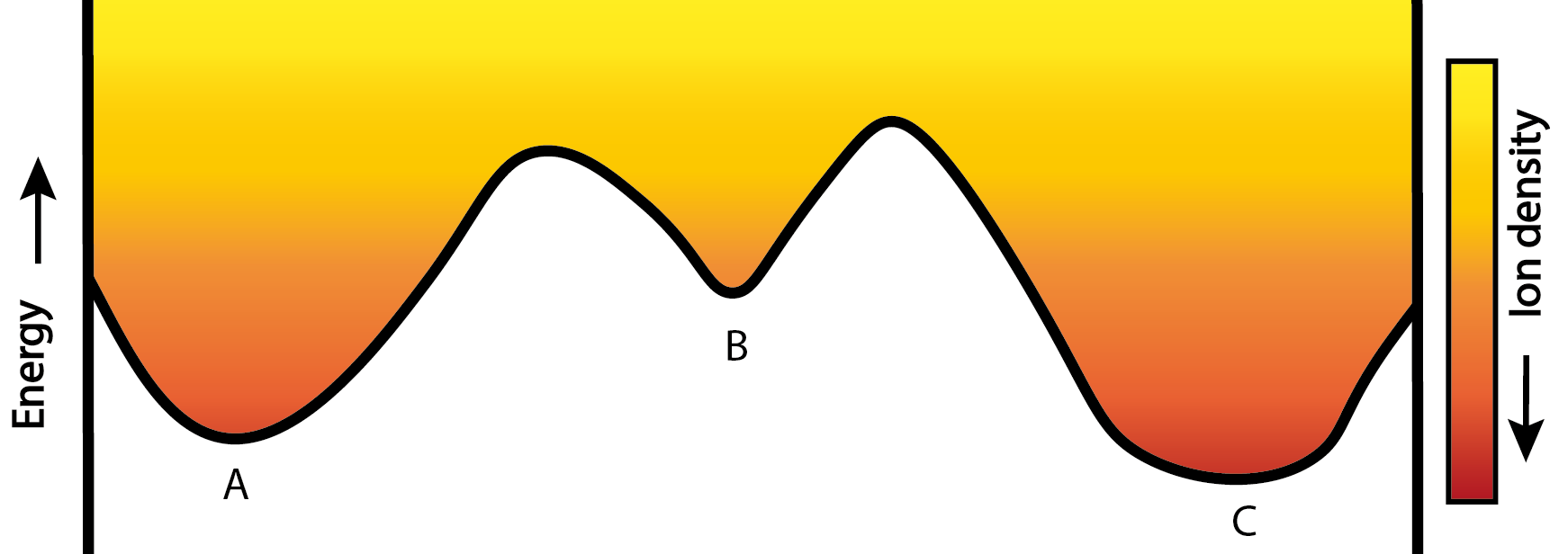}	
		\caption{Energy landscape and the corresponding ion density, with site A, B, and C}
		\label{fig:e_landscape}
	\end{center}
\end{figure}
As shown in Figure \ref{fig:e_landscape} a jump from site A to B can have a different energy barrier as the reverse jump due to the difference in site energy, even though the number of A-B and B-A jumps will be the same in equilibrium.
By taking into account the residence time at each site an 'effective' jump rate can be determined, thus taking the effect of the site energy into account.
The activation energy can then be determined by comparing the ratio between the effective jump rate and the attempt frequency. The effective jump rate, $\Gamma_\mathrm{eff}$, differs from the jump rate in equation \ref{eq:gamma} by taking into account the fraction of time that the diffusing atoms occupies a site ($o$):
\begin{equation}
	\Gamma_\mathrm{eff} = \frac{J}{N t o}
	\label{eq:eff_jumprate}
\end{equation}
Using the effective jump rate the activation energy ($\Delta E_{A}$) can be calculated \cite{Vineyard_1957}: 
\begin{equation}
	\Delta E_{A} = -k_\mathrm{B} T \ln(\frac{\Gamma_\mathrm{eff}}{\nu^{*}})
	\label{eq:e_act}
\end{equation}
where $k_B$ is Boltzmann's constant, $T$ the temperature in Kelvin, and $\nu^{*}$ the attempt frequency.
Since jump and attempt frequencies are temperature dependent the activation energy may also be a function of temperature. 
Such non-Arrhenius behaviour can be investigated by performing MD simulations at different temperatures. 
Finally, the difference in activation energy between back and forth jumps provides the energy difference between two sites, which can be used to predict changing site-occupancies with temperature.

\subsection{Collective jumps}
The large concentration of diffusing atoms in solid electrolytes, 31 mol/L in \ce{\beta-Li3PS4}, is likely to result in interactions between the diffusing Li-atoms. This potentially causes collective jump processes \cite{Alex_hollandite}, which may have a severe impact on macroscopic diffusion \cite{He_2017, Xu_2012}. 
MD simulations are a powerful tool for investigating complicated collective jump processes \footnote{The process which we call 'collective' jumps is also referred to as 'correlated' or 'concerted' jumps in the literature.}. 
Knowing the position and time of jumps allows to determine if jumps are correlated in time and space. 
At present the spatial condition for correlated motion is assumed to be slightly larger as the largest jump distance between Li-sites in the crystal. For \ce{\beta-Li3PS4} this is 4.5 \AA~, in which 4 \AA~ between the bc-planes is the largest jump distance. 
Because each diffusing atom changes direction after a vibration it is unlikely that collective motions last for multiple vibrations. A rational time condition for collective motion thus is the average time of a single vibration, which is equal to the period of the attempt frequency ($\frac{1}{\nu^{*}}$ seconds).
Clearly the conditions that define transitions as collective are debatable and should be chosen carefully for each material. 

\subsection{Radial Distribution Functions}
The atomic environment determines the forces and energy barriers that govern the behaviour of diffusing atoms. 
The atomic environment can be represented using a Radial Distribution Function (RDF), which effectively reveals the density of different elements as a function of distance from the atom of interest. 
For example, this has shown to be useful for \ce{Na3PS4}, where the RDF's from MD simulations suggested that Na-vacancies are essential for increased Na-diffusion \cite{Klerk_Na}. 
As the position of all atoms is available at any time step from a MD simulation, RDF's can be readily obtained for any site, atom or element.

\subsection{Summary}
Summarising, if the crystallographic sites are known, detailed diffusion properties can be extracted from MD simulations. In principle a single MD simulation, in which each type of jump occurs a significant number of times, already provides detailed insight into diffusion, including the diffusion path, attempt frequency, jump rates, activation energies, collective motions, atomic environments, and the correlation factor. \\
For a thorough understanding MD simulations at several temperatures might be necessary, for instance in the case of non-Arrhenius behaviour, or to investigate the reliability of the results over a range of temperatures.
Extracting the described information is beneficial for understanding of the diffusion process,  allowing for a targeted approach to design and prepare materials with enhanced properties, as will be demonstrated in this study for the Li-ion conductor \ce{\beta-Li3PS4}.   

\section{Example: \ce{\beta-Li3PS4}}
\ce{Li3PS4} has been a well-known Li-ion conductor since the 1980's \cite{Mercier_1982}, but interest grew after experiments with nano-sized crystals showed a Li-ion conductivity of $1.6*10^{-4}$ S/cm \cite{Liu_2013}, approaching the value that is required for solid state Li-ion batteries.
Three polymorphs of \ce{Li3PS4} have been reported \cite{Homma_2011}, the low-temperature $\gamma$-phase, the $\beta$-phase at intermediate temperatures, and the high temperature $\alpha$-phase. The $\beta$-phase shows the highest room temperature conductivity of the three polymorphs \cite{Liu_2013}, and is thus most interesting for application as a solid electrolyte. 
A beneficial property of \ce{\beta-Li3PS4} is its apparent stability against Li-metal \cite{Liu_2013}, although DFT-calculations report otherwise \cite{Richards_2016, Zhu_2015}.
\ce{Li3PS4} can be prepared via a solvent route \cite{Nguyen_2016, Teragawa_2014}, resulting in a conductivity of 3.3*10$^{-4}$ S/cm \cite{Nguyen_2016}, enabling coating of cathode materials. In this way no additional solid electrolyte material needs to be added in the cathodic mixture \cite{Teragawa_2014}, resulting in a larger effective energy density in combination with a small interface resistance. \\
Several studies \cite{Mercier_1982, Homma_2011, Chen_2015} investigating the structure of \ce{\beta-Li3PS4} report significantly different Li-ion positions and occupancies. Neutron diffraction \cite{Chen_2015} indicates that the coordinates of the Li-ion 4c position strongly depend on temperature, potentially explaining the differences between X-ray diffraction studies \cite{Mercier_1982, Homma_2011}. \\
Based on the larger sensitivity to Li-ions of neutrons compared to X-rays, the Li-positions determined from neutron diffraction \cite{Chen_2015} at 413 K are used for the analysis of the present MD simulations on \ce{\beta-Li3PS4}.

\subsection{Effect of Li-vacancies and Li-interstitials}
The introduction of Li-vacancies has been suggested to be beneficial for Li-ion conductivity in \ce{\beta-Li3PS4} \cite{Phani_2017}, whereas the high ionic conductivity of the isostructural \cite{Nishino_2014} compound \ce{Li10GeP2S12} (= \ce{Li_{3.33}Ge_{0.33}P_{0.67}S4}) suggests that introducing extra Li-ions in \ce{\beta-Li3PS4} can also lead to an increased Li-ion conductivity. To study the effect of both Li-vacancies and Li-interstitials on the diffusion mechanism DFT MD simulations were performed for \ce{\beta-Li3PS4}, \ce{\beta-Li_{2.75}PS4}, and \ce{\beta-Li_{3.25}PS4} at 450, 600 and 750 K. 

\subsubsection{Li-ion diffusion}
\begin{figure}[tbhp]
	\begin{center}
		\begin{subfigure}{0.4\textwidth}
			\includegraphics[width=\textwidth]{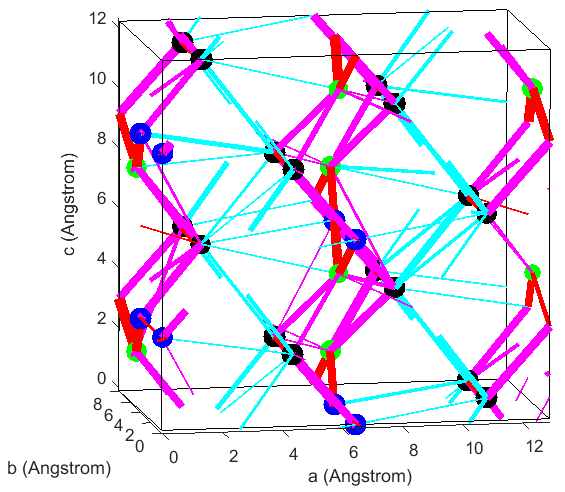}	
			\caption{~}
			\label{fig:paths_vac}
		\end{subfigure}
		\begin{subfigure}{0.4\textwidth}
			\includegraphics[width=\textwidth]{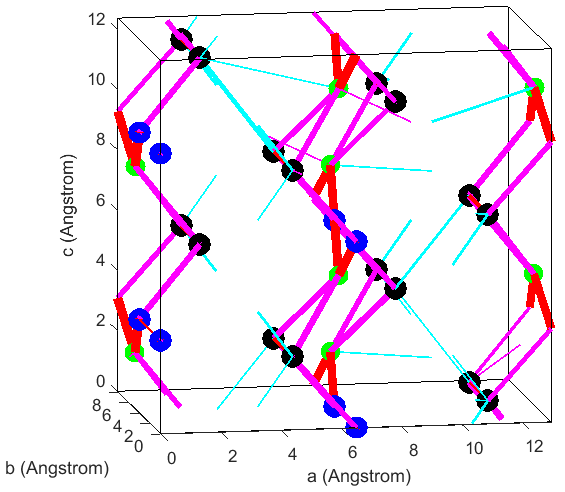}	
			\caption{~}
			\label{fig:paths_normal}
		\end{subfigure}
		\begin{subfigure}{0.4\textwidth}
			\includegraphics[width=\textwidth]{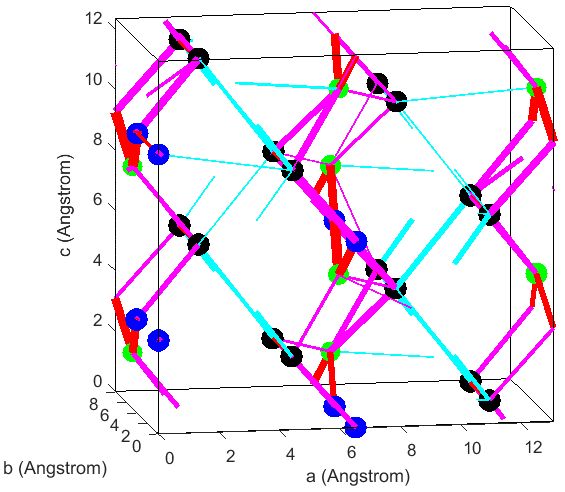}	
			\caption{~}
			\label{fig:paths_inter}
		\end{subfigure}
	\caption{Jump diffusion paths at 600 K for (a) \ce{\beta-Li_{2.75}PS4}, (b) \ce{\beta-Li3PS4}, and (c) \ce{\beta-Li_{3.25}PS4}. Li-ion sites are shown by: 4b = blue, 4c = green, and 8d = black. The primary direction of jumps is shown by: a-axis = cyan, b-axis = red, and c-axis = pink, thicker lines correspond to larger jump rates.}
	\label{fig:paths_all}
	\end{center}
\end{figure}

\begin{figure}[htbp]
	\begin{center}
		\includegraphics[width=0.45\textwidth]{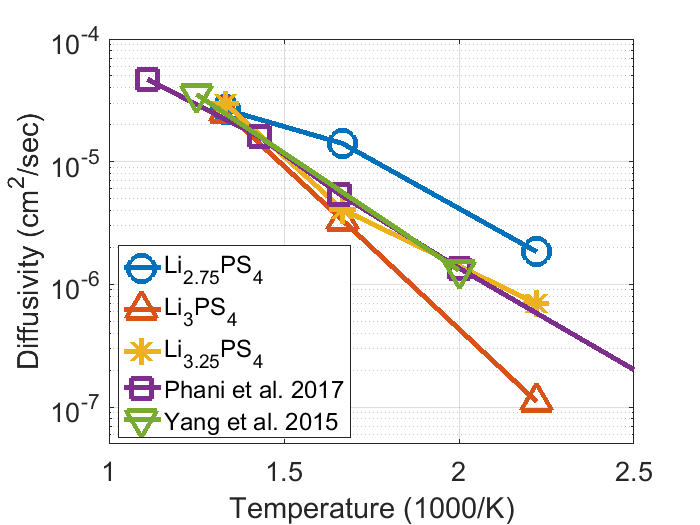}				
		\caption{Tracer diffusivity from the current MD simulations, Phani et al. \cite{Phani_2017} and Yang et al. \cite{Yang_2015}}
		\label{fig:tracer_vac_inter}
	\end{center}
\end{figure}

The diffusion paths from simulations at 600 K, shown in Figure \ref{fig:paths_all}, demonstrate that diffusion takes place along the b-axis via 4b-4c jumps, along the c-axis via 4b-8d and 4c-8d jumps, and through interplane 8d-8d jumps in the a-direction. 
In stoichiometric \ce{\beta-Li3PS4} relatively few transitions occur between the bc-planes, indicating that Li-ion diffusion occurs primarily within the bc-planes. 
When Li-vacancies or Li-interstitials are introduced the Li-ion diffusion within the bc-plane remains similar to the stoichiometric composition \ce{\beta-Li3PS4}, while the amount of jumps between bc-planes increases significantly, resulting in three-dimensional diffusion. \\
The beneficial effect of the three-dimensional diffusion is reflected in the tracer diffusivity, shown in Figure \ref{fig:tracer_vac_inter}. The Li-ion diffusivity in \ce{\beta-Li_{2.75}PS4} is almost an order of magnitude larger than in \ce{\beta-Li3PS4}. Introducing Li-interstitials by creating \ce{\beta-Li_{3.25}PS4} also results in a larger diffusivity, especially at the lowest simulated temperature. \\
Based on the tracer diffusivity the conductivity of \ce{\beta-Li3PS4} is $10^{-2}$ S/cm at 450 K, comparable to impedance experiments \cite{Nguyen_2016} at the same temperature. 
Extrapolating the Li-ion diffusivity of \ce{\beta-Li3PS4} to 110 $^\circ$C results in a Li-ion diffusivity of 1*10$^{-8}$ cm$^2$/sec, close to the values reported by NMR experiments: $3.0*10^{-8}$ cm$^2$/sec at 100 $^\circ$C \cite{Gobet_2014} and between $10^{-6}$ and $10^{-8}$ cm$^2$/sec at 120 $^\circ$C \cite{Hayamizu_2016}. \\
The results from the current MD simulations on \ce{\beta-Li3PS4} are comparable to the values reported previously \cite{Phani_2017, Yang_2015}, except at 450 K. This anomaly is most likely caused by the shorter simulation times of the previous studies, which can lead to an overestimation of the tracer diffusion at low temperatures. 
At 750 K all the MD simulations show a similar value for the diffusivity, which can be explained by the melted lithium sub-lattice \cite{Phani_2017} at this temperature. After melting the lithium ordering over the different crystallographic sites disappears, which seems to have a larger impact as the deviating stoichiometries investigated here.

\subsubsection{Jump rates}
\begin{figure}[htbp]
	\begin{center}
		\includegraphics[width=0.45\textwidth]{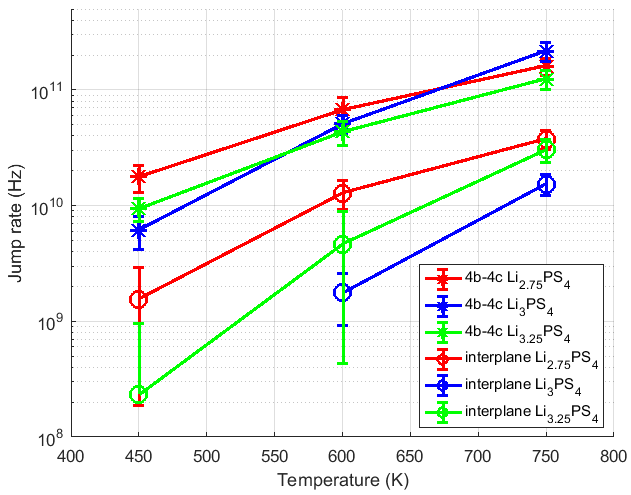}	
		\caption{Jump rates for the 4b-4c and interplane jumps}
		\label{fig:jump_rates}
	\end{center}
\end{figure}

The differences in tracer diffusivities between the three compositions can be explained by the rate-limiting jump mechanism. 
The most frequent jump process is the 4b-4c transition, the rate of which is comparable between the three compositions, as shown in Figure \ref{fig:jump_rates}. However, to obtain three-dimensional diffusion paths in \ce{\beta-Li3PS4} interplane jumps are necessary, the rate of which is significantly different for the three compositions, also shown in Figure \ref{fig:jump_rates}. With lower temperature these differences increase, and in \ce{\beta-Li3PS4} at 450 K no interplane jumps occurred. \\ 
Because two-dimensional diffusion processes have a smaller correlation factor compared to three-dimensional processes \cite{Mehrer_book, Friauf_1962} the tracer diffusivity in \ce{\beta-Li3PS4} is significantly lower, even though the jump rate of the fastest diffusion process is similar. 

\subsubsection{Activation energies}
\begin{figure}[tbhp]
	\begin{center}
		\begin{subfigure}{0.45\textwidth}
			\includegraphics[width=\textwidth]{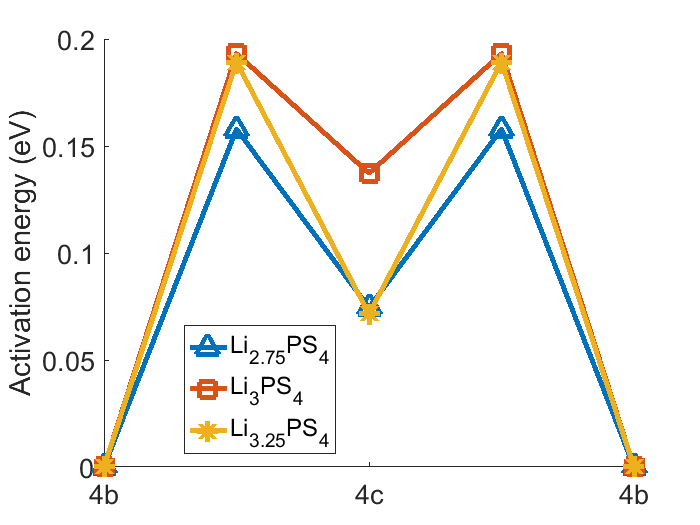}	
			\caption{~}
			\label{fig:b_barriers}
		\end{subfigure}
		\begin{subfigure}{0.45\textwidth}
			\includegraphics[width=\textwidth]{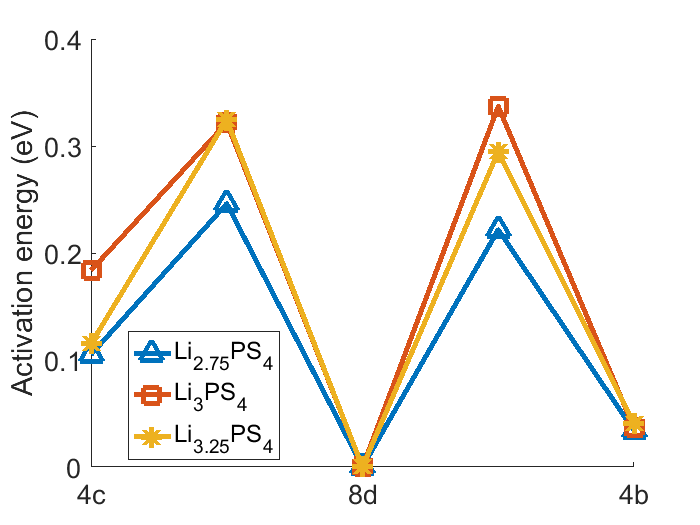}	
			\caption{~}
			\label{fig:c_barriers}
		\end{subfigure}
		\caption{Activation energy at 600 K along (a) the b-axis, and (b) the c-axis.}
		\label{fig:all_barriers}
	\end{center}
\end{figure}

The activation barriers for diffusion along the b- and c-axis obtained from the MD simulations are shown in Figure \ref{fig:all_barriers}. At 600 K the interplane 8d-8d jumps show activation energies of 0.41 eV for \ce{\beta-Li3PS4}, 0.35 eV for \ce{\beta-Li_{3.25}PS4}, and 0.28 eV for \ce{\beta-Li_{2.75}PS4}. 
It should be noted that other jump processes also occur in the simulations, however, their significantly larger activation energy indicates that these will not contribute significantly to Li-ion diffusion and are therefore left out of the current analysis. \\
Over the simulated temperature range the activation energies for \ce{\beta-Li_{2.75}PS4} are constant. For the other two compositions the same holds for jumps starting at the 4c- and 8d-sites. However, jumps starting at 4b-sites show a decreasing activation energy with increasing temperature, indicating that the 4b-sites become unstable at high temperature. \\
To validate the activation energies from MD simulations comparison with experimental values is important, however, a wide distribution in values is reported based on electrochemical experiments: 0.16 \cite{Homma_2011}, 0.32 \cite{Nguyen_2016}, 0.36 \cite{Liu_2013}, and 0.47 \cite{Teragawa_2014} eV. 
NMR experiments resulted in activation energies of 0.40 eV for macroscopic diffusion and 0.09 eV for local jumps \cite{Hayamizu_2016}. Given this wide distribution of values a comparison of experimental activation energies with the present simulations seems unreasonable. \\
Simulations also report a wide range of activation energies. NEB calculations on \ce{\beta-Li3PS4} report activation energies of 0.3 eV along the a-axis and 0.2 eV along the b- and c-axis \cite{Lepley_2013}, while other NEB calculations \cite{Yang_2016} report 0.26 eV along the a- and b-axis and 0.08 eV for collective Li-ion jumps in the b-direction, and bond-valence calculations report values of 1.0 eV along the a-axis and 0.8 eV in the bc-plane \cite{Xiao_2015}. 
The results from NEB calculations and MD simulations are comparable, while bond-valence calculations appear to overestimate the activation energy. 
The activation energies from MD simulations indicate that diffusion along the b-axis is most facile, followed by diffusion along the c-axis, and along the a-axis diffusion is most difficult, in agreement with results from neutron diffraction \cite{Chen_2015}.

\subsubsection{Collective jump processes}
\begin{figure}[tbhp]
	\begin{center}
		\begin{subfigure}{0.3\textwidth}
			\includegraphics[width=\textwidth]{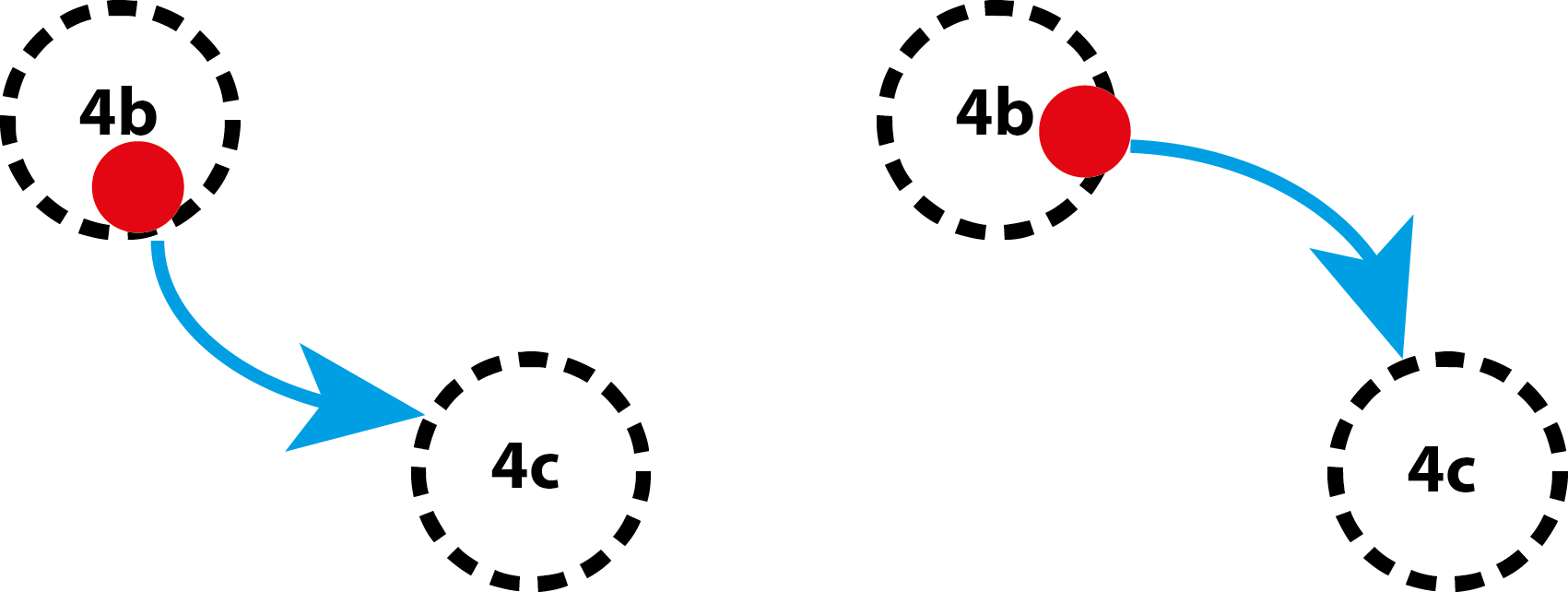}	
			\caption{~}
			\hrule
		\end{subfigure}
		\begin{subfigure}{0.3\textwidth}
			\includegraphics[width=\textwidth]{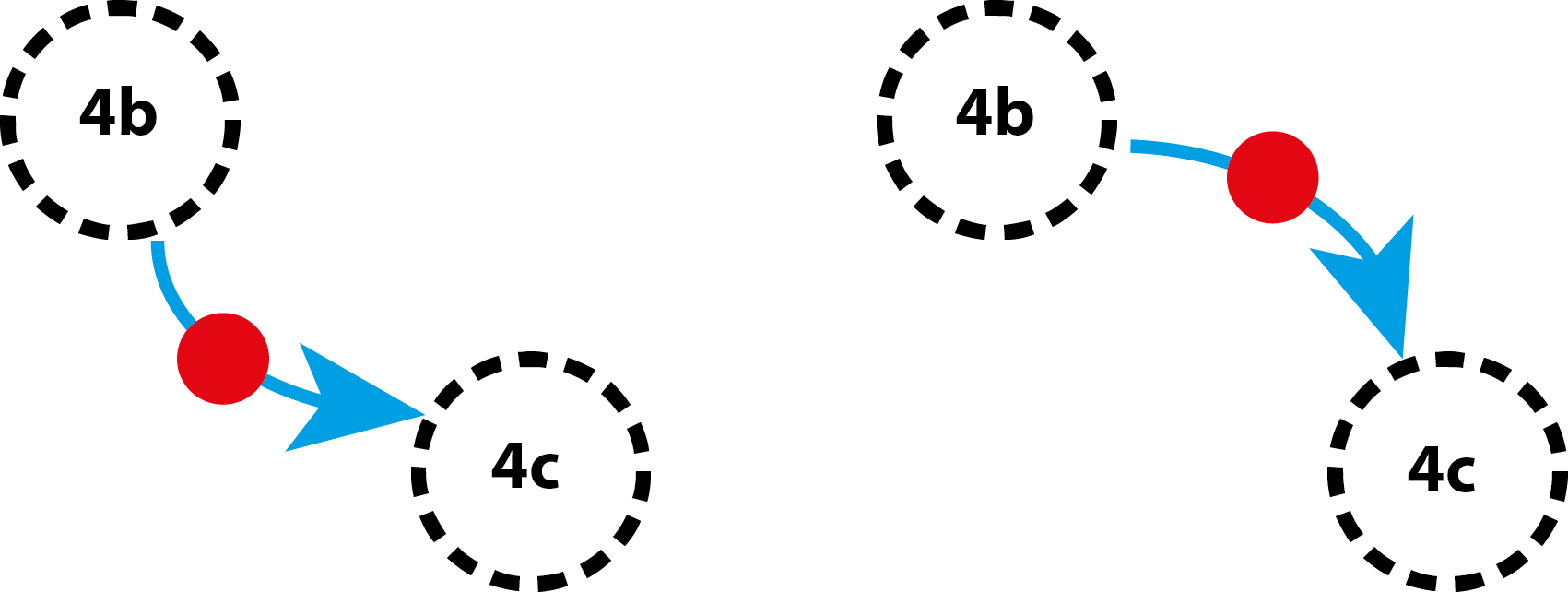}	
			\caption{~}
			\hrule
		\end{subfigure}
		\begin{subfigure}{0.3\textwidth}
			\includegraphics[width=\textwidth]{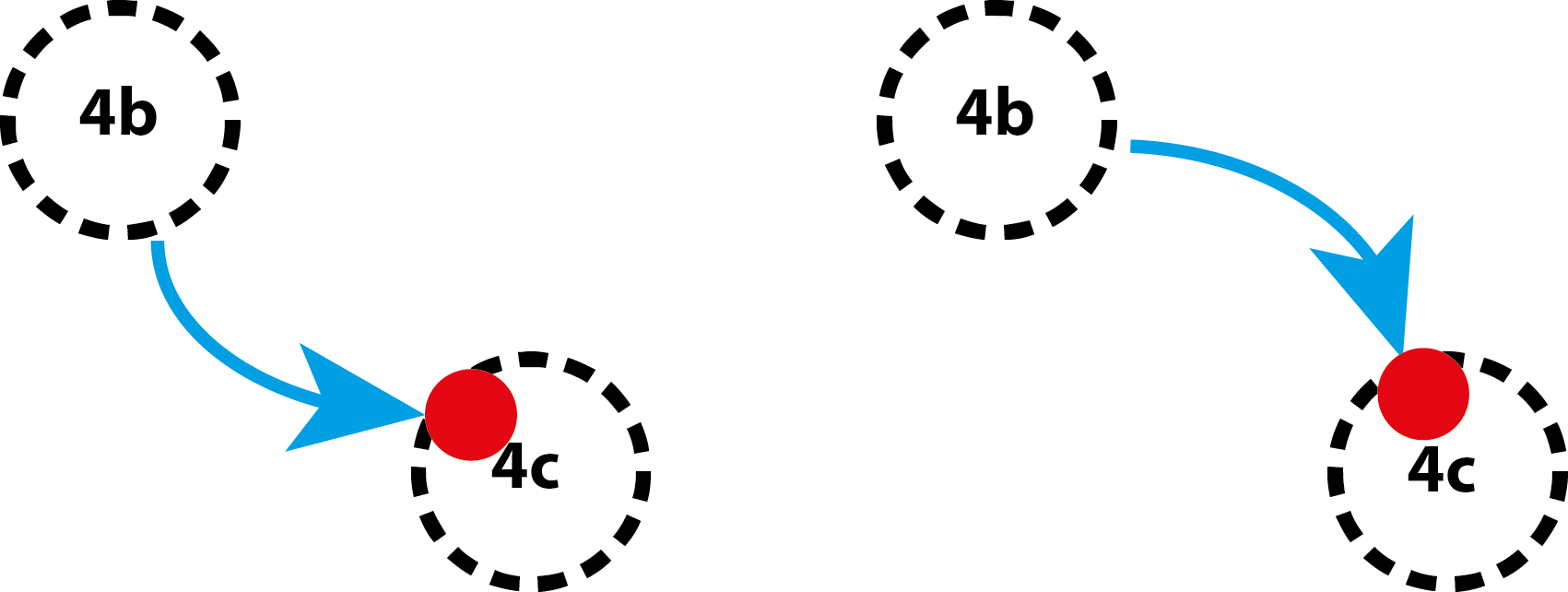}	
			\caption{~}
		\end{subfigure}
	\caption{Schematic picture of Li-ion (red spheres) motion during a collective 4b-4c transition along the b-axis. (a) At the start of, (b) during, and (c) after the transition.}
	\label{fig:scheme_collective}
	\end{center}
\end{figure}

Given the large lithium concentration of 31 mol/L in \ce{\beta-Li3PS4} Li-ions can be expected to interact strongly with each other. Yang et al. \cite{Yang_2016} reported the presence of collective jumps in \ce{\beta-Li3PS4} by showing that the activation energy for diffusion along the b-axis is just 0.08 eV for two Li-ions moving collectively, while it is 0.26 eV for a single Li-ion. \\
\begin{figure}[htbp]
	\begin{center}
		\includegraphics[width=0.45\textwidth]{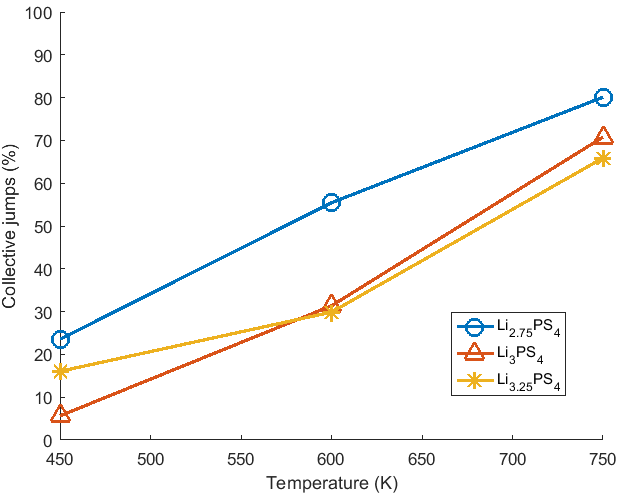}	
		\caption{Percentage of collective jumps in the MD simulations}
		\label{fig:collective}
	\end{center}
\end{figure}
Analysis of collective jumps in the present MD simulations reveals that the percentage of collective jumps depends on the temperature and Li-concentration, as shown in Figure \ref{fig:collective}. 
The percentage of collective jumps displays a strong increase with temperature, where 65 to 80 \% of the jumps occur collectively at 750 K. Although at 450 K the simulations show less collective jumps, still 24 \% of the jumps is collective in \ce{Li_{2.75}PS4}. 
The large percentages indicate that collective jump processes may have a significant effect on the Li-ion diffusion in \ce{\beta-Li3PS4}, especially at elevated temperatures. However, it should be noted that this analysis strongly depends on the conditions specified for collective jumps, which in this case are that two jumps occur within one period of the attempt frequency and are less than 4.5 \AA~ apart.\\ 
Further analysis shows that in \ce{\beta-Li3PS4} the collective jumps are primarily simultaneous 4c-4b jumps and simultaneous 4b-4c jumps, schematically shown in Figure \ref{fig:scheme_collective}. 
In \ce{Li_{3.25}PS4} collective 4b-4c and 4c-4b jumps also occur most frequently, additionally 4b-8d jumps collective with 4b-4c jumps occur, and interplane jumps are often collective with 4b-4c jumps. 
In the simulations of \ce{Li_{2.75}PS4} different collective behaviour is observed, where the combination of 4b-8d jumps and 4b-4c jumps is occurring most frequently. Collective 4b-4c jumps also occur frequently, but significantly less compared to the other compositions. \\ 
Collective jumps involving more than two Li-ions also occur in the current MD simulations, in some cases involving up to 5 atoms. The collective movement of multiple atoms is complex and difficult to analyse. However, it should be anticipated that collective motion of several ions induces large ionic conductivities, as observed in \ce{Li10GeP2S12} \cite{Xu_2012}, making this an interesting subject for further investigation. 

\subsubsection{Attempt frequency}
\begin{figure}[htbp]
	\begin{center}
		\begin{subfigure}{0.45\textwidth}
			\includegraphics[width=\textwidth]{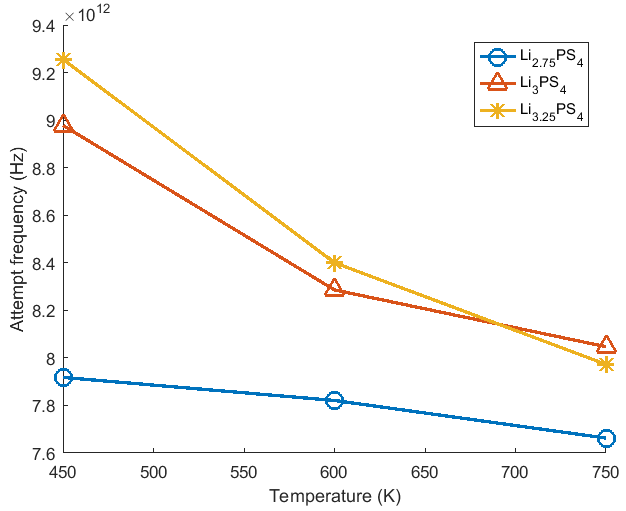}	
			\caption{~}
			\label{fig:attempt_freqs}
		\end{subfigure}			
		\begin{subfigure}{0.45\textwidth}
			\includegraphics[width=\textwidth]{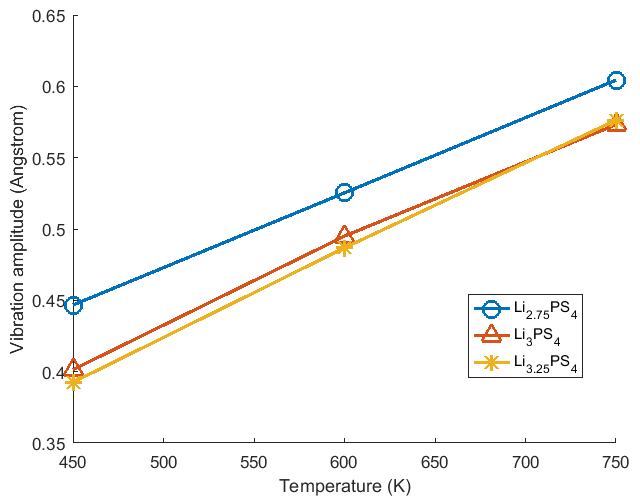}	
			\caption{~}
			\label{fig:vib_amplitudes}			
		\end{subfigure}
		\caption{(a) Attempt frequencies and (b) vibration amplitudes from the MD-simulations.}
	\end{center}
\end{figure}

The attempt frequencies obtained from the simulations are shown in Figure \ref{fig:attempt_freqs}, in all simulations the attempt frequency is close to the typically assumed $10^{13}$ Hz \cite{vdVen_2001, Koettgen_2017}. The present analysis shows that a straightforward Fourier transformation of the ionic velocity from a MD simulation can be used to determine the attempt frequency. \\
Figure \ref{fig:attempt_freqs} demonstrates that the attempt frequency decreases with increasing temperature, this decrease with increasing temperature can be understood by the change in the vibrational amplitude, shown in Figure \ref{fig:vib_amplitudes}. For \ce{Li_{3.25}PS4} and \ce{Li3PS4} the vibrational amplitude increases by approximately 35\% between 450 and 750 K, while the speed of the atoms increases by just 29\% (based on: $E = \frac{1}{2} m v^2$). The average vibration time therefore increases, leading to a decreasing attempt frequency. \\
This example demonstrates that there can be a significant temperature dependence on the attempt frequency, which may go unnoticed using the more regular static computational methods. 
Furthermore, the attempt frequency and vibration amplitude in \ce{Li_{2.75}PS4} differ by approximately 10\% from the other simulations, showing that relatively small changes in the crystal structure can have a significant effect on these properties. 
Thereby the present simulations and analysis indicate that consideration of the attempt frequency and its dependence on structure and temperature is of significant importance in quantifying and understanding ionic diffusion. 

\subsection{Effects of doping}
The MD simulations on \ce{\beta-Li_{3.25}PS4} show that creating Li-interstitials in \ce{\beta-Li3PS4} induces three dimensional diffusion, increasing the macroscopic Li-ion conductivity, in line with experimental work \cite{Nishino_2014}. 
Li-vacancies also induce three dimensional diffusion, and the MD simulations indicate significantly larger Li-ion conductivity compared to the introduction of Li-interstitials. However, we are unaware of work which has explored the impact of Li-vacancies on Li-ion diffusion in \ce{\beta-Li3PS4}. 
To determine the impact of introducing Li-vacancies by doping MD simulations were performed on \ce{\beta-Li_{2.75}PS_{3.75}Br_{0.25}}, in which two S-atoms were replaced by Br-atoms. \\
Additionally, the impact of oxygen doping is investigated, since this is also reported as a strategy to improve the Li-ion diffusivity \cite{Xiao_2015, Gobet_2014}.
However, O-doping does not change the Li-content, and the mechanism of higher Li-ion diffusivity by O-doping has not been revealed. To gain understanding of how oxygen doping increases the Li-ion diffusivity MD simulations on \ce{\beta-Li3PS_{3.75}O_{0.25}} were performed by replacing two S-atoms by O-atoms.

\subsubsection{Br-doping}
\begin{figure}[tbhp]
	\begin{center}
		\begin{subfigure}{0.45\textwidth}
			\includegraphics[width=\textwidth]{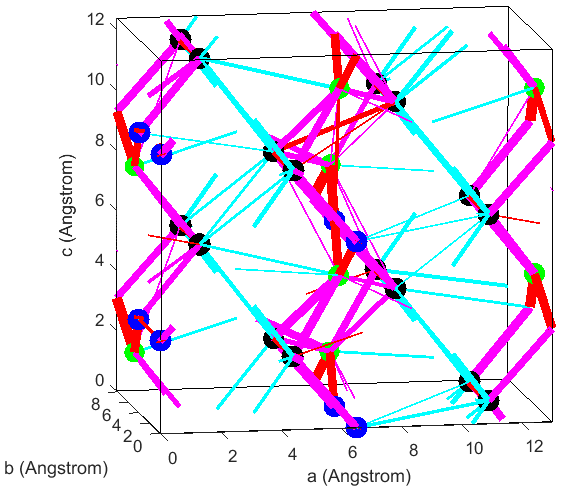}	
			\caption{~}
			\label{fig:paths_bromide}
		\end{subfigure}
		\begin{subfigure}{0.45\textwidth}
			\includegraphics[width=\textwidth]{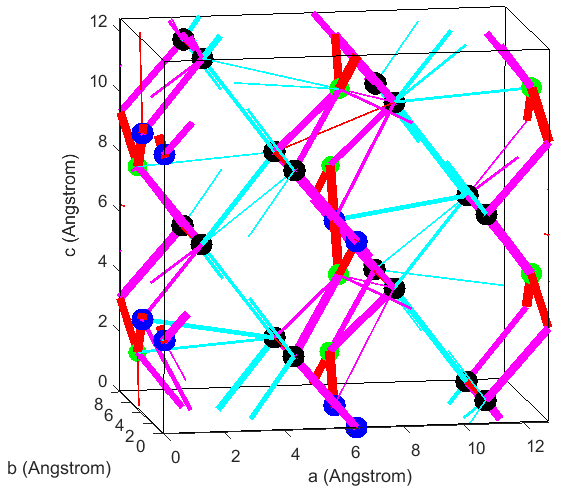}	
			\caption{~}
			\label{fig:paths_oxygen}
		\end{subfigure}
	\caption{Jump diffusion paths at 600 K for (a) \ce{\beta-Li_{2.75}PS_{3.75}Br_{0.25}} and (b) \ce{\beta-Li3PS_{3.75}O_{0.25}}. Li-ion sites are shown by: 4b = blue, 4c = green, and 8d = black. The primary direction of jumps is shown by: a-axis = cyan, b-axis = red, and c-axis = pink, thicker lines correspond to larger jump rates.}
	\label{fig:paths_doped}
	\end{center}
\end{figure}

\begin{figure}[tbhp]
	\begin{center}
		\begin{subfigure}{0.45\textwidth}
			\includegraphics[width=\textwidth]{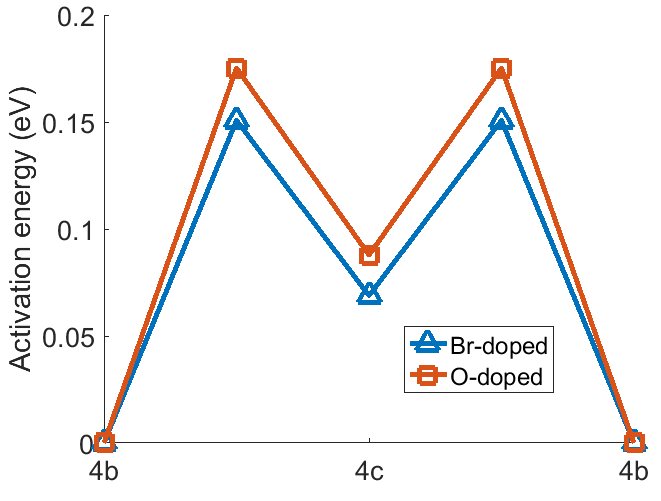}	
			\caption{~}
			\label{fig:barrier_b_doped}
		\end{subfigure}
		\begin{subfigure}{0.45\textwidth}
			\includegraphics[width=\textwidth]{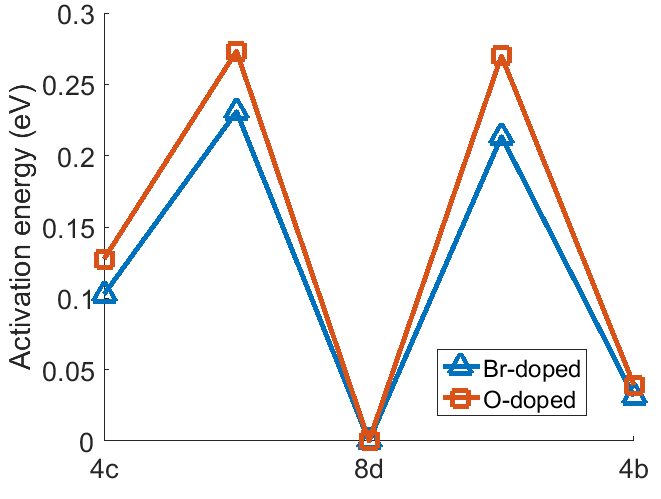}	
			\caption{~}
			\label{fig:barrier_c_doped}
		\end{subfigure}
		\caption{Activation energies at 600 K for the Br- and O-doped materials along (a) the b-axis, and (b) the c-axis.}
	\label{fig:barriers_doped}
	\end{center}
\end{figure}

The jump diffusion path from the MD-simulation of \ce{\beta-Li_{2.75}PS_{3.75}Br_{0.25}} is shown in Figure \ref{fig:paths_bromide}. As should be anticipated from the simulations of \ce{\beta-Li_{2.75}PS4}, Figure \ref{fig:paths_bromide} shows that Br-doping leads to three-dimensional diffusion paths.
In \ce{\beta-Li_{2.75}PS_{3.75}Br_{0.25}} the tracer diffusivity results in 1.56*10$^{-6}$, 1.01*10$^{-5}$, and 3.71*10$^{-5}$ cm$^2$/sec at 450, 600 and 750 K, respectively, comparable to the diffusivities of \ce{\beta-Li_{2.75}PS4}. 
The activation energies for diffusion along the b- and c-axis in the Br-doped composition at 600 K, shown in Figure \ref{fig:barriers_doped}, differ by just 0.02 eV from the MD simulation with Li-vacancies. The activation energy for interplane jumps is 0.28 eV in both \ce{\beta-Li_{2.75}PS_{3.75}Br_{0.25}} and \ce{\beta-Li_{2.75}PS_{4}}. \\
The similar activation energies in \ce{\beta-Li_{2.75}PS_{4}} and \ce{\beta-Li_{2.75}PS_{3.75}Br_{0.25}} indicates that the primary cause of the high Li-ion diffusivity in Br-doped \ce{\beta-Li3PS4} are the Li-vacancies. For other dopants which introduce Li-vacancies similar results are thus expected, suggesting that there are many ways of increasing the Li-ion diffusivity in \ce{\beta-Li3PS4}. 

\subsubsection{O-doping}
\begin{figure}[htbp]
	\begin{center}
		\includegraphics[width=0.45\textwidth]{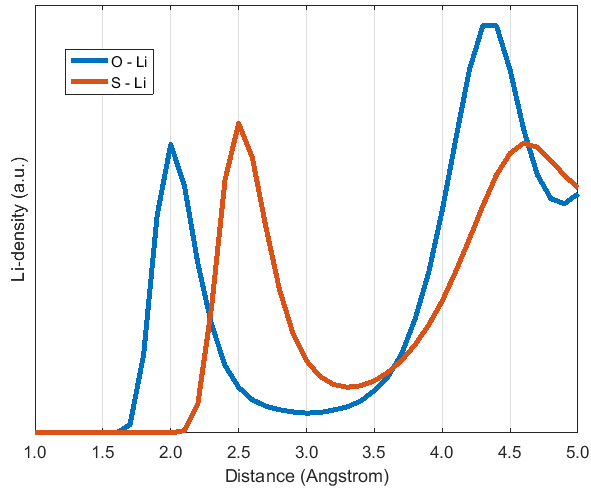}	
		\caption{O-Li and S-Li distribution in \ce{\beta-Li_{3}PS_{3.75}O_{0.25}} at 600 K.}
		\label{fig:rdf_diff_o}
	\end{center}
\end{figure}

The jump diffusion paths from the present MD-simulations on \ce{\beta-Li3PS_{3.75}O_{0.25}}, shown in Figure \ref{fig:paths_oxygen}, demonstrate that O-doping also induces three-dimensional diffusion, as predicted by Xiao et al. \cite{Xiao_2015}.
Compared to \ce{\beta-Li3PS4} doping with oxygen leads to a larger Li-ion diffusivity, and tracer diffusivities of 6.96*10$^{-7}$, 7.59*10$^{-6}$, and 2.52*10$^{-5}$ cm$^2$/sec at 450, 600 and 750 K, respectively. 
The introduction of oxygen results in smaller activation energies compared to \ce{\beta-Li3PS4}, as shown in Figure \ref{fig:barriers_doped}. The biggest impact is observed for the interplane jumps, which have an activation energy of just 0.32 eV in \ce{\beta-Li3PS_{3.75}O_{0.25}}, comparable to the Li-rich \ce{\beta-Li_{3.25}PS4}. \\
This is surprising because the introduction of oxygen does not affect the Li-concentration.  
To investigate this the radial distribution functions (RDF's) for oxygen and sulphur in \ce{\beta-Li_{3}PS_{3.75}O_{0.25}} are shown in Figure \ref{fig:rdf_diff_o}.
The smaller ionic radius of oxygen \cite{Shannon_1976} results in a smaller O-Li distance compared to the S-Li distance.  
However, the Li-density of the first coordination shell in the RDF is significantly lower around the O-atoms. Integrating the Li-density up to 3.5 \AA~ shows that (on average) there are 2.9 Li-atoms in the first coordination shell of O-atoms, and 3.5 Li-atoms in the first coordination shell of S-atoms. \\
This implies that it is unfavourable for Li-ions to be near the O-atoms in \ce{\beta-Li3PS_{3.75}O_{0.25}}, and these Li-ions must be accommodated elsewhere within the crystal structure. Oxygen-doping thus creates Li-vacancies near the O-atoms and Li-interstitials elsewhere, explaining why O-doping has a beneficial effect on Li-ion diffusivity in \ce{\beta-Li_{3}PS_{4}}. \\
It is usually assumed that a higher polarisability leads to higher diffusivity \cite{Bachman_2016} via lower activation energies caused by lattice softening \cite{Kraft_2017}. In the case \ce{\beta-Li3PS_{3.75}O_{0.25}} the higher diffusivity caused by O-atoms, which have lower polarisability compared to S-atoms, demonstrates that the opposite can also occur. 
The RDF's shown in Figure \ref{fig:rdf_diff_o} indicate that the site-energy near the O-atoms is higher, which, if the transitions state energy stays the same, lowers the activation energy. In this case the less polarisable O-atoms are thus beneficial for Li-ion diffusion.

\section{Conclusions}
The present approach demonstrates that from a single MD simulation key properties for ionic diffusion can be obtained, through which a thorough understanding of diffusion can be developed. 
The example of DFT based MD simulations on \ce{\beta-Li3PS4} indicates that Li-ion jumps between bc-layers are the slowest jump process, limiting the macroscopic conductivity. Adding Li-interstitials or Li-vacancies significantly promotes these transitions, initiating three-dimensional diffusion, which enhances the Li-ion diffusivity.
Li-vacancies can be introduced through Br-doping at the sulphur sites, which is predicted to result in an order of magnitude larger Li-ion conductivity in \ce{\beta-Li3PS4}. Furthermore, it is shown that oxygen-doping at the sulphur site induces the formation of local interstitials, rationalizing the increased Li-ion diffusivity that has been reported.
The thorough analysis of MD simulations presented is a general approach that can be applied to all crystalline ionic conductors, which can help to build understanding of diffusional processes in solid state electrolytes, and provide direction to the design of improved solid electrolyte materials.
The Matlab-code developed for the analysis of MD simulations is freely available online \cite{Bitbucket}. 

\section{DFT calculations}
The DFT simulations were performed using VASP \cite{Kresse_1993}, using the GGA approximation \cite{Perdew_1996} and the PAW-PBE basis set \cite{Blochl_1994}. A cut-off energy of 400 eV was used for simulations containing oxygen, and 280 eV for the other simulations. 
The \ce{\beta-Li3PS4}-phase crystallises in the orthorhombic space-group Pnma (no. 62), with lattice parameters of a = 12.82, b = 8.22, and c = 6.12 \AA~ at 637 K \cite{Homma_2011}. The crystal structure as measured by Homma et al. \cite{Homma_2011} was used as a starting point for the structure minimisations. To prevent self-interactions a 1x1x2 super cell was used in the calculations. 
After minimisation without symmetry restrictions the lattice angles were close to 90$^{\circ}$ in all cases, and the lattice parameters changed by less as 2\%, with the a- and c-parameters showing a small increase, while the b-parameter decreased slightly. 
During the minimisations a k-point mesh of 4x6x4 was used, which was reduced to a k-point mesh of 1x2x1 for the MD simulations. 
The total simulation time of the MD simulations was 500 ps., with 2 fs. time-steps. The first 2.5 ps. were used as equilibration time and were thus not used for the analysis. Simulations were performed in the NVT ensemble, with temperature scaling after every 50 time-steps. \\

\section{Acknowledgements}
Financial support from the Advanced Dutch Energy Materials (ADEM) program of the Dutch Ministry of Economic Affairs, Agriculture and Innovation is gratefully acknowledged. The research leading to these results has received funding from the European Research Council under the European Union's Seventh Framework Programme (FP/2007-2013)/ERC Grant Agreement nr. [307161] of M.W. The authors would like to thank Tomas Verhallen and Casper Versteylen for fruitful discussions over many cups of coffee and Alexandros Vasileiadis for testing the Matlab code.

\bibliography{bib_md_analysis}

\end{document}